\documentclass[conference]{IEEEtran}
\usepackage[hyphens]{url}
\usepackage{hyperref}

\usepackage[backend=biber, style=ieee]{biblatex}
\usepackage{amsmath,amssymb,amsfonts}
\usepackage{algorithmic}
\usepackage{graphicx}
\usepackage{textcomp}
\usepackage{xcolor}
 \usepackage{balance}

\usepackage{listings}
\usepackage{tabularray}

\usepackage{geometry}
\begin{document}

\date{}

\title{\Large \bf SETC: A Vulnerability Telemetry Collection Framework}

\author{\IEEEauthorblockN{Ryan Holeman}
\IEEEauthorblockA{\textit{Beacom College} \\
\textit{Dakota State University}\\
Madison, SD, USA \\
ryan.holeman@trojans.dsu.edu}
\and
\IEEEauthorblockN{John Hastings}
\IEEEauthorblockA{\textit{Beacom College} \\
\textit{Dakota State University}\\
Madison, SD, USA \\
john.hastings@dsu.edu}
\and
\IEEEauthorblockN{Varghese Mathew Vaidyan}
\IEEEauthorblockA{\textit{Beacom College} \\
\textit{Dakota State University}\\
Madison, SD, USA \\
varghese.vaidyan@dsu.edu}
}

\maketitle


\begin{abstract}

  As emerging software vulnerabilities continuously threaten enterprises and Internet services, there is a critical need for improved security research capabilities. This paper introduces the Security Exploit Telemetry Collection (SETC) framework - an automated framework to generate reproducible vulnerability exploit data at scale for robust defensive security research. SETC deploys configurable environments to execute and record rich telemetry of vulnerability exploits within isolated containers. Exploits, vulnerable services, monitoring tools, and logging pipelines are defined via modular JSON configurations and deployed on demand. Compared to current manual processes, SETC enables automated, customizable, and repeatable vulnerability testing to produce diverse security telemetry. This research enables scalable exploit data generation to drive innovations in threat modeling, detection methods, analysis techniques, and remediation strategies. The capabilities of the framework are demonstrated through an example scenario. By addressing key barriers in security data generation, SETC represents a valuable platform to support impactful vulnerability and defensive security research.
\end{abstract}


\section{Introduction}

Annually, there is a consistent increase in the identification of new software vulnerabilities that pose threats to enterprises and Internet services \cite{qualsys2023}. In order to investigate, understand, and research these vulnerabilities, security practitioners and researchers rely on data produced during exploitation of these vulnerabilities. This data is commonly sourced from real-world incidents or simulated environments. However, in both cases, the process for collection and the data gathered is limited.

In cases where data is obtained from the wild or sanctioned events, such as collegiate defense competitions, data could be better in many ways. One of the most significant limitations is that the production of this data is not repeatable. If a researcher requires extra telemetry, data points, or sources that are not recorded, there is no practical way to obtain these data points. Secondly, researchers are restricted to events created outside of their control. If researchers want to study a specific vulnerability or attack technique, they can not control what attackers choose to exploit.

Researchers are often bound by time and resource conditions when data is obtained through attack simulation. In order to manually reproduce an attack, researchers must create a vulnerable target, recreate an attack, and log their activity. More tooling is needed to support the automation around attack simulation that supports rich logging and customization capabilities.

This research introduces Security Exploit Telemetry Collection (SETC), a framework to remedy these limitations and provide a feature-rich system for producing this data. The SETC framework records deep security telemetry of attacks against vulnerable services. The framework achieves the collection of security telemetry by hosting and exploiting vulnerable services in a controlled container environment. Vulnerable services, exploits, and telemetry collection are modular and defined in framework configuration files. Compared to current manual processes, SETC enables automated, customizable, and repeatable vulnerability testing to produce diverse security telemetry. This research enables scalable exploit data generation to drive innovations in threat modeling, detection methods, analysis techniques, and remediation strategies. 

This paper is organized as follows. Section \ref{background} reviews the background and motivating factors for the creation of SETC. Section \ref{overview} provides an overview of SETC's capabilities, followed by details of the system design of the framework in Section \ref{design}.  Section \ref{example} shows an example of the usage of SETC with a sample set of vulnerable systems, exploits, and telemetry sources. Section \ref{limitations} discusses the limitations and shortcomings of the current framework. Section \ref{future} presents ideas for future work and SETC roadmaps. Section \ref{related} introduces related work, followed by the conclusion in Section \ref{conclusion}.

\section{Background \& Motivation}\label{background}

The primary motivation of this research is to create a platform that provides rich quantitative data for attack and vulnerability research. Having a uniform platform that provides exploit automation, customizable environments, extensible logging, and repeatable scenarios will expand, accelerate, and reinforce various areas of defensive security research. There are many research platforms for offensive security research, such as Metasploit, Empire, and Canvas. However, there have yet to be any comparable platforms focused on defensive security research.

Currently, researchers have access to solutions that partially achieve some capabilities of SETC. However, these existing solutions are either limited or immature. Many of them concentrate on establishing education-centric environments \cite{RN107}, lab environments, or cyber ranges \cite{RN108, RN134}. Unfortunately, these systems require extensive manual effort to achieve automated end-to-end exploitation and data generation. While a handful of systems do prioritize end-to-end exploit setup and automation, their focus is often confined to specific vulnerability classes, such as PHP \cite{RN132} or local operating system vulnerabilities. Suffice it to say that these systems lack the comprehensive approach and scope that both current and forthcoming versions of SETC aim to provide.\footnote{More details on these related tools appears in Section \ref{related}.}

The first goal for SETC is to provide a straightforward method to automate end-to-end vulnerability exploitation. The framework utilizes containers to host both vulnerable systems and systems capable of performing exploits against these vulnerable hosts. SETC employs container frameworks such as Docker and Kubernetes to deploy these systems. To enhance user-friendliness, the setup and deployment of these systems are abstracted into simple JSON-based configuration files.

In order to produce a modular, configurable, and repeatable architecture, SETC utilizes modern container and orchestration tools. The framework deploys numerous containers and sidecar containers \cite{RN133} dedicated to producing and collecting various forms of telemetry. These auxiliary containers can also be used to deploy custom tooling such as endpoint agents, network taps, and system monitors.

SETC is also designed to incorporate an end-to-end logging pipeline. This allows for collecting logs at various sources and storing logs in systems such as SIEMs (security information and event management), databases, or simple file stores. This framework component provides a method for exporting and sharing data from system runs. The logging pipeline also plays a crucial role in data transformation so that logs can be converted to various logging standards or formats. As SIEMs can be deployed and populated in SETC sessions, the framework can train detections, test alerts, or provide a platform for general log analysis.

SETC creates a foundation for various forms of future research. The data it produces can be used to test and validate popular logging standard formats such as CIM (Common Information Model) \cite{RN20}, OCSF \cite{RN14}, and CEF \cite{RN19}. It can be used to test proactive and reactive security tools. It can also be used for general vulnerability research, such as vulnerability classification, detection methods, and exploit analysis. Section \ref{future} provides some insight into some future work being considered with the use of the framework.

\section{System Overview}\label{overview}

\begin{figure*}[!ht]
    \centering
    \includegraphics[width=7in]{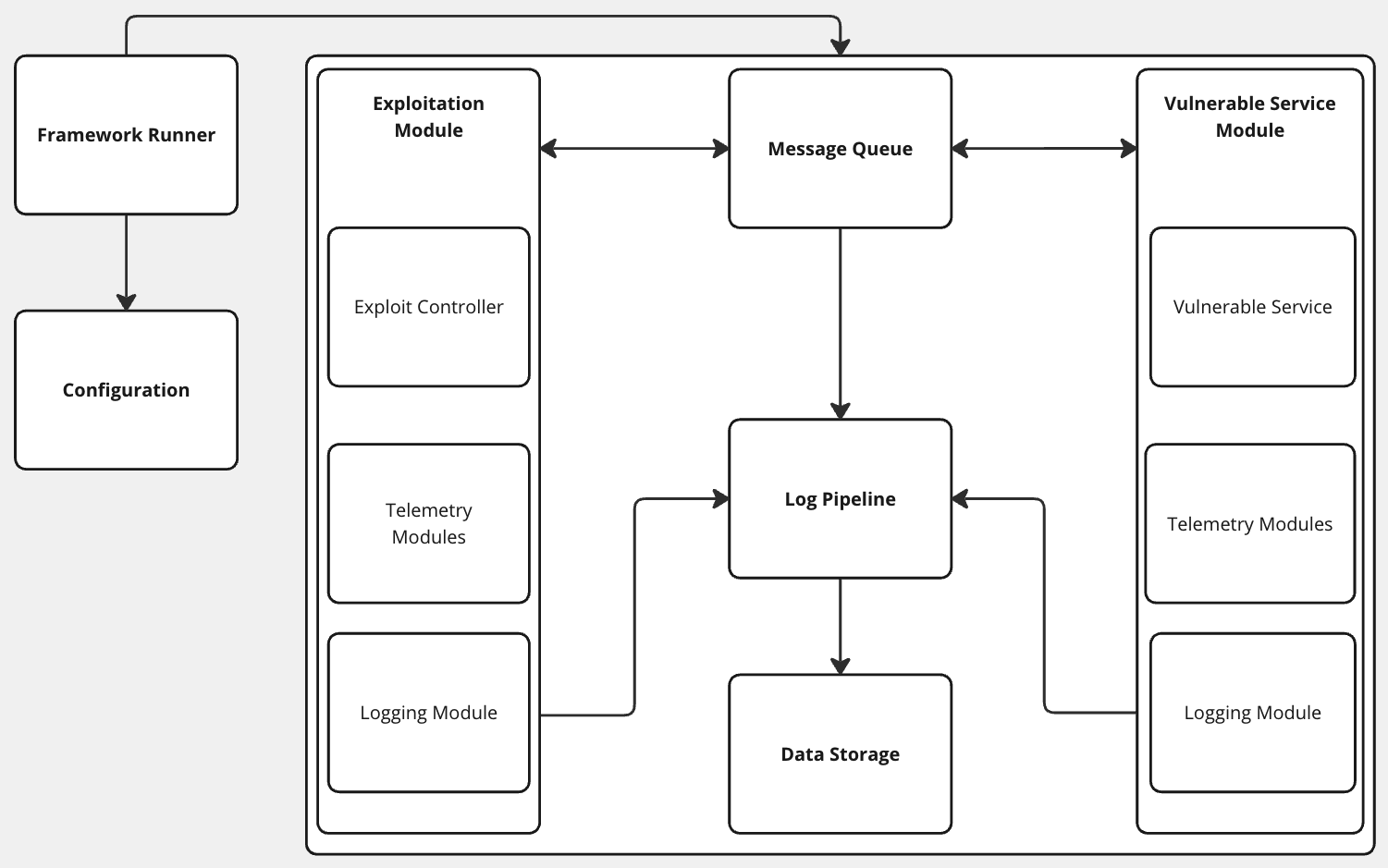}
    \hfil
    \caption{SETC framework design.}
    \label{fig:design}
\end{figure*}

In its current version, the core of SETC is a Docker API application, Docker files, and a JSON configuration file. Before running the framework, a user must create or choose an existing configuration file. This configuration file defines vulnerable systems that will be created, exploited, and monitored. Upon initiating the framework, the core framework runner will parse and read a configuration file that dictates how a particular execution of the framework will behave. Each entry in the configuration file instructs what vulnerable service should be hosted and what exploit should be run against it. These groups of entries may define classes of vulnerabilities, vulnerabilities associated with specific software, or vulnerabilities from particular date ranges.

Once a configuration is parsed, the framework will initialize the vulnerable service instance containers in a private network. The framework will parallelly initiate the telemetry collection modules for each vulnerable instance. The collection modules are a mix of proxy container services and container sidecars with various security metric collection capabilities. Once the framework has validated vulnerable services and collection modules are fully running, the framework will transition into exploitation.

At the start of each exploitation phase, the framework will create containers capable of running end-to-end exploits specific to a vulnerable instance. Once an exploit is initiated, the framework will monitor the exploit containers for signs of successful or failed exploitation. In the event of a failed exploit attempt, the framework will reinitiate an attack and repeat until a successful exploit is achieved. After completing an exploit, the framework will transition into a clean-up and telemetry collection phase.

During the telemetry collection phase, data is sent to a logging pipeline. The logging pipeline of the framework serves two core purposes. The first purpose is to function as a data transposition layer for log events in the pipeline. Telemetry files can be converted into various logging standard formats. These include standards such as OCSF, CIM, and CEF. Secondly, the logging pipeline phase routes data to its final destination after data transposition. These destinations are configurable and include sinks such as simple file storage or SIEM ingestion and analysis.

\section{System Design}\label{design}
SETC is an automated system that creates a containerized infrastructure to execute and record evidence of remote system exploitation. SETC orchestration is handled by a Python application that utilizes Docker API. This orchestration script, illustrated as the framework runner in Figure \ref{fig:design}, is the heart of the framework. It manages and maintains all system hosts, actions, and data production.

SETC requires various Docker containers to function. The Docker containers used by the framework fall into three functional categories: vulnerable systems, attacker systems, and auxiliary systems. Containers for each category have expected behavior they should follow.

\begin{enumerate}
    \item \textbf{Vulnerable Systems} - Container instances hosting specific vulnerabilities.
    \item \textbf{Attacker Systems} - Container instances capable of exploiting specific vulnerabilities.
    \item \textbf{Auxiliary Systems} - Container instances providing services for logging, telemetry collection, or other SETC functionality.
\end{enumerate}

Vulnerable system containers are simply hosts vulnerable to a specific attack or exploit. The vulnerable services or software in these containers are expected to be fully configured and initiated upon start with no manual intervention. Aside from these simple criteria, there are no other requirements for vulnerable systems. This allows for the use of vulnerable containers from various sources that provide vulnerable Docker images, such as capture the flags (CTFs), proof of concept source code repositories, and image-sharing services. Many containers, images, and Dockerfiles used as vulnerable systems in  SETC are sourced from Vulhub \cite{RN33}, open-source security research repositories, Dockerhub, and software vendors.

Attacker system containers are hosts that contain a tool, script, or application capable of exploiting a vulnerability hosted on a vulnerable system. Attacker systems currently support two different workflows. For supported offensive security toolsets, containers such as metasploitframework/metasploit can be run in vanilla mode and passed the Metasploit exploit module name as a configuration parameter. This allows a single attacker image to be used on various vulnerable systems. This method also provides many shortcuts in the SETC configuration schema to reduce the complexity of configuration elements. Alternatively, attacker systems can provide a startup script that contains the command line execution of a tool, script, or application for running an exploit against a vulnerable system. This type of workflow is helpful for various proof-of-concept exploit scripts found in security research proof-of-concept repositories.

Auxiliary systems comprise the rest of SETC. These systems provide telemetry sources, log pipelines, data storage, and any other tool or service desired for the SETC environment. Security tooling, monitoring, and telemetry sources should be deployed as sidecar containers. This ensures they will be associated with specific networks, vulnerable or attack systems. Auxiliary systems that do not run as sidecars are less ephemeral and persist throughout a SETC session. These systems provide logging pipelines, data transposition services, data collection, and analysis.

\begin{figure}
    \centering
    \lstset{breaklines=true, frame=single, 
    linewidth=0.98\linewidth}
    \begin{lstlisting}[%label=fig_log,
        basicstyle=\small]
{"name":"CVE-2018-11776",
"settings": {
"description":"Struts2 OGNL injection RCE",
"target_image":"vulhub/struts2:2.5.25",
"attack_src":"msf",
"exploit":"multi/http/struts2_multi_eval_ognl"
}},
    
\end{lstlisting}
    \caption{Example SETC configuration entry.}
    \label{fig:configuration}
\end{figure}

SETC configuration files glue all of these systems together. They define what vulnerabilities will be tested, what telemetry will be collected, and how data will be output and stored. They allow SETC to be highly modular and customizable. The configuration format was designed to be simple and compact, allowing for extensive configurability when needed. An example configuration entry is provided in Figure \ref{fig:configuration}. This example showcases condensed configuration syntax, which is made possible through SETC configuration shortcuts for supported attack systems.

The system is designed to be useful for small or large-volume vulnerability research. Vulnerabilities can be run in parallel for quick analysis on commodity hardware. To provide a better understanding of all of the framework modules and how they interact, the system's core components are described below and illustrated in Figure \ref{fig:design}.

\begin{enumerate}
    \item \textbf{Framework Runner} - The framework runner is the core component of the framework. It is an application that automatically generates multi-container Docker applications based on a configuration file. These multi-container environments are designed to create vulnerable systems, exploit these systems, and log telemetry from various event sources.
    \item \textbf{Configuration} - Specialized framework configuration files to define various aspects of the multi-container systems. The configuration can specify what exploits are to be run, what information should be logged, what tools should be installed, and what logging standards should be used.
    \item \textbf{Exploitation Module} - These containers run automated exploitation against defined vulnerable services in the multi-container network.
    \item \textbf{Vulnerable Service Module} - These containers run known vulnerable services that can be exploited.
    \item \textbf{Sub-modules} - Sub-modules can exist on any container in the system. These commonly install telemetry and logging tools that integrate with the logging pipeline.
    \item \textbf{Message Queue} - Facilitates state between exploitation modules and vulnerable service modules.
    \item \textbf{Logging Pipeline} - The logging pipeline collects logs and telemetry from each container in the system. The logging pipeline can also be used to transpose data into various formats.
    \item \textbf{Data Storage} - The data storage container stores all data sent to the logging pipeline. By default, this system persists stored data after a framework runner job for further analysis.
\end{enumerate}

\section{Tests \& Evaluation}\label{example}
In order to demonstrate the capabilities and use of SETC, we will construct a simple example that produces data to solve a research question. For this example, SETC will produce data that can be used to answer the following scenario: ``Given a sample set of HTTP vulnerabilities, can the CIM logging standard be used to identify signatures of exploitation for all vulnerabilities in the set?" This example, posed as a user story, showcases how SETC is run, the data it can produce, and how that data can be used to answer the example question.

\begin{table}
\label{tab_vulns}
\renewcommand{\arraystretch}{1.3}

\caption{Vulnerabilities included in the example SETC configuration file}
\label{tab:vuln_containers}
\centering
\begin{tabular}{|c|p{0.8in}|c|}
\hline
CVE & Description & Container Source\\
\hline
CVE-2018-11776 & Struts2 OGNL injection RCE & Vulhub\\
\hline
CVE-2005-2877 & Twiki history RCE & Metasploitable\\
\hline
CVE-2021-3129 & Laravel debug RCE & Vulhub\\
\hline
CVE-2021-42013 & HTTP apache normalize RCE & SETC\\
\hline
CVE-2021-41773 & HTTP apache normalize RCE & SETC\\
\hline
CVE-2021-25646 & Apache druid JS RCE & Vulhub\\

\hline
\end{tabular}
\end{table}

To start, we will select a sample set of HTTP vulnerabilities. For simplicity, this test will use six vulnerabilities from the example configuration file provided in the SETC project. As the scenario focuses on HTTP vulnerabilities, the six selected vulnerabilities will only contain web-based vulnerabilities. Table \ref{tab:vuln_containers} lists the vulnerabilities, exploits, and the source where the vulnerable container files were sourced.

For this example, the SETC logging pipeline is configured to produce CIM HTTP logs. SETC has also been configured to output logging data in a Docker shared volume and deploy a Splunk instance. The data can be analyzed directly from flat files in the shared volume or the Splunk instance.

Upon initiation of SETC for this configuration file, the following steps happen for each defined vulnerability:
\begin{enumerate}
    \item A private network is set up that will contain the vulnerable service container, attacker container, and any container sidecars.
    \item A container instance is started that hosts a server vulnerable to the specified CVE.
    \item A container instance is started that contains tools to exploit the vulnerable system.
    \item Monitoring sidecars that will record telemetry for this environment are deployed.
    \item An exploit is initiated from the attacker container against the vulnerable container.
    \item SETC monitors the environment to confirm that an exploit and its attack activity have been completed.
    \item Logging telemetry is flushed to the logging pipeline outside of the exploit environment.
    \item All containers and private networks are shut down and destroyed.
\end{enumerate}

\begin{figure}[!t]
    \centering
    \includegraphics[width=.75\linewidth]{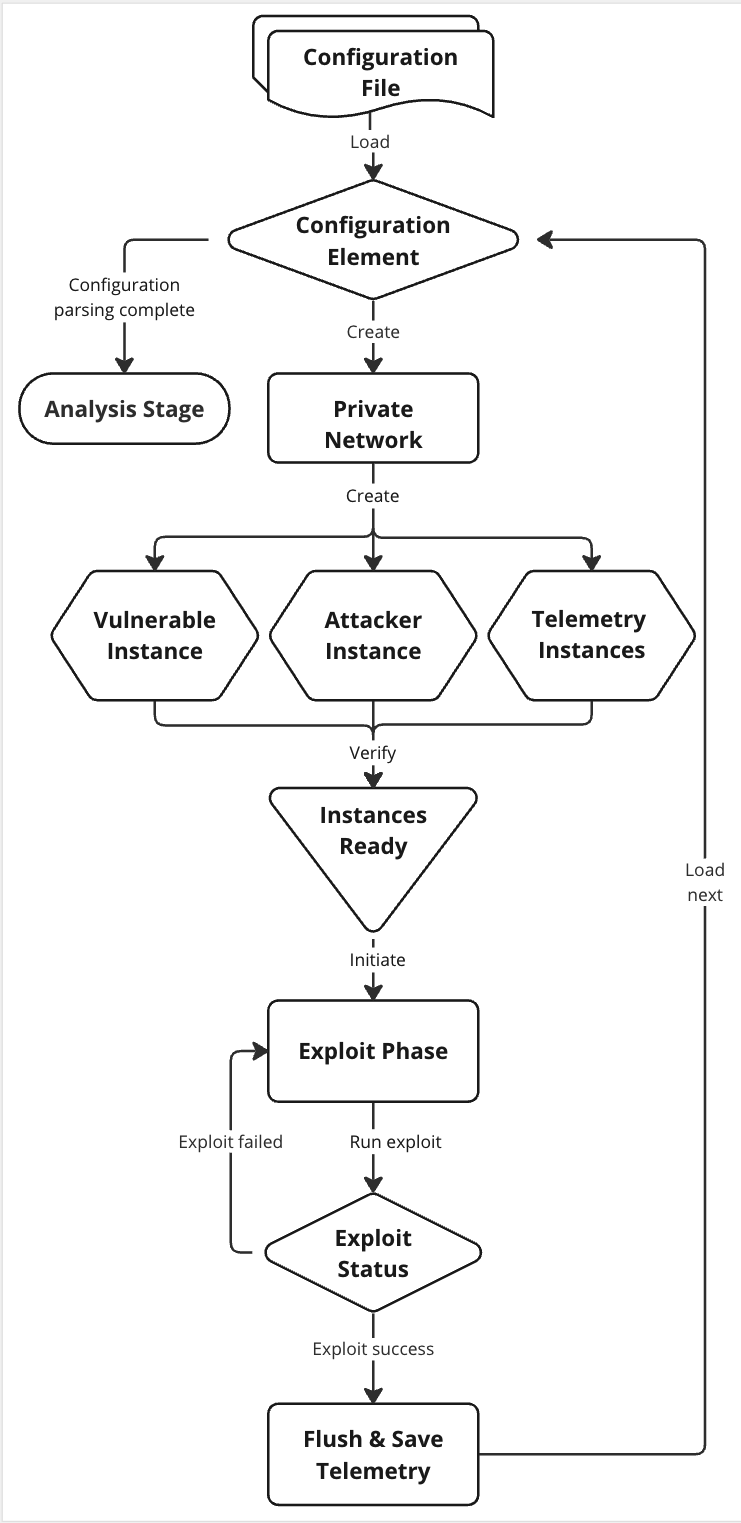}
    \caption{SETC workflow.}
    \label{fig:workflow}
\end{figure}

These steps are then repeated for all entries defined in the SETC configuration file. As Splunk is configured to run in this instance, the Splunk container is left running and attached to the session logging volume. This allows for immediate analysis of the data produced by the session. The complete workflow is illustrated in Figure \ref{fig:workflow}.

This scenario is small and only contains six vulnerabilities, so a manual analysis is conducted on the CIM HTTP logs produced in Splunk. For context on the brevity of the logs, the session of six exploited vulnerabilities only produces 16 log events in total for all vulnerable containers. A sample CIM log entry from this test is shown in Figure \ref{fig:log_sample}.

\begin{figure}[!t]
    \centering
    \lstset{breaklines=true, frame=single, linewidth=0.98\linewidth}
    \begin{lstlisting}[%label=fig_log,
        basicstyle=\small]
{'timestamp': 1690410131.023336, 'action': 'http', 'bytes': 429, 'bytes_in': 429, 'bytes_out': 0, 'category': '-', 'dest': '172.29.0.3', 'dest_port': 80, 'http_content_type': ['text/plain'], 'http_method': 'tkHe', 'http_referrer': '-', 'http_referrer_domain': '-', 'http_user_agent': 'Mozilla/5.0 (Macintosh; Intel Mac OS X 12_2_1) AppleWebKit/605.1.15 (KHTML, like Gecko) Version/15.2 Safari/605.1.15', 'http_user_agent_length': 116, 'host': 'target_CVE-2021-42013', 'src': '172.29.0.4', 'status': 200, 'uri_path': '/cgi-bin/.2e/.2e/.2e/.2e/.2e/bin/sh', 'url': '/cgi-bin/.2e/.2e/.2e/.2e/.2e/bin/sh', 'url_length': 35}
    
\end{lstlisting}
    \caption{Example CIM log entry.}
    \label{fig:log_sample}
\end{figure}

Based on a manual analysis of the logs produced, we can determine that exploit signatures were only present in CIM HTTP logs for three of the six vulnerabilities examined. This analysis is shown in Table \ref{tab:vun_sigs}. While advanced signature techniques involving fields such as bytes in, bytes out, user agent, HTTP method, and others are possible, this example only considers obvious exploit detection based on payload string matching.

\begin{table}[!t]
\renewcommand{\arraystretch}{1.3}
\caption{Signature analysis results for SETC example}
\label{tab:vun_sigs}
\centering
\begin{tabular}{|c|p{0.8in}|c|}
\hline
CVE & Description & Signature in Logs \\
\hline
CVE-2018-11776 & Struts2 OGNL injection RCE & False \\
\hline
CVE-2005-2877 & Twiki history RCE & True \\
\hline
CVE-2021-3129 & Laravel debug RCE & False \\
\hline
CVE-2021-42013 & HTTP apache normalize RCE & True \\
\hline
CVE-2021-41773 & HTTP apache normalize RCE & True \\
\hline
CVE-2021-25646 & Apache druid JS RCE & False \\

\hline
\end{tabular}
\end{table}

Three vulnerabilities were not detected in the example CIM HTTP logs. This is because the exploit payload is contained within POST data. As with many logging standards, POST data is not included as a standard field because it typically contains sensitive or private data. While full or partial POST data can be included in customized or extended logging standards, this topic is outside the scope of this example scenario.

Finally, it is essential to point out that this scenario is repeatable. Using SETC to rerun this configuration file would produce the same results. While log files from consecutive runs may contain temporal changes or variations due to exploit behavior, the core activity in the logs should be identical. This aspect of SETC provides a mechanism for easily sharing data, reproducing data, and testing scenarios with new tools or telemetry components. In the case above, if we want to extend the CIM HTTP model to include POST data, we could use the same SETC configuration file to test and validate our logging enhancements.

\section{Limitations}\label{limitations}
The most significant limitations of SETC stem from the design choice to use Docker and Docker API as its underlying container engine. This design choice limits sidecar capabilities, multi-system container provisioning, and container operating system support. These Docker sidecar limitations restrict the use of sidecars for endpoint-specific tool deployment and telemetry collection. With SETC's current reliance on Docker, only network-based sidecars are supported. SETC is also limited to single-system Docker files. Multi-system orchestration files such as Dockercompose are not supported due to their limited support by Docker API. Finally, using Docker limits the ability to use Windows-based containers for vulnerability testing. Work is underway to transition SETC to a Kubernetes engine. This transition will resolve these identified technical limitations and significantly expand SETC's capabilities.

Currently, SETC focuses on remotely accessible vulnerabilities. While local vulnerability exploitation and monitoring can be achieved through exploit chaining, the framework runner workflow and configuration schema are not optimized for this scenario. Direct support for local vulnerability testing environments will have to be built in future releases of SETC.

Finally, there are external limitations that impact SETC. As pointed out by \citeauthor*{RN135} \cite{RN135} in their research paper on web application security, there is a lack of exploits available compared to the number of existing CVEs. This imbalance limits test data for research or requires time-consuming exploit development work. Similarly, during the development of SETC, it was discovered that there is a lack of generally available vulnerable container images for available exploits. While sources such as Vulhub \cite{RN33} provide preconfigured vulnerable instances for known exploits, the amount of available and working containers is limited.

\section{Future Work}\label{future}
As eluded above, SETC has many technical limitations due to its reliance on Docker and Docker API. Current work is underway to port the underlying SETC container engine to Kubernetes. While SETC, in its current state, is fully functional and highly useful for vulnerability research, its current version is intended to serve as a proof of concept. As such, the current version of SETC is considered to be in the alpha stage. After completing the work to transition SETC to Kubernetes, the release will mark a stable beta version for the project.

Work is also underway to investigate the capabilities and gaps of popular logging standards concerning vulnerability exploits and attacks. Popular logging standards, including OCSF, CIM, and CEF, are being analyzed with SETC to understand what fields are most significant for signature detection and what underlying signature data points are missing from event fields. This work aims to improve current logging standards and identify which ones provide the most benefits and comprehensive telemetry.

Finally, an initial investigation is ongoing into how SETC can improve current and newly proposed security detection capabilities. The data SETC was designed to produce can aid and test signature and anomaly-based security detections. It can also be used to create and test the validity of new detection techniques, such as ML-based approaches and providence graph security alerts \cite{RN53, RN54}.

\section{Related Work}\label{related}
Work related to SETC falls into three areas. The most prominent types of research related to SETC are frameworks that provide end-to-end automated exploitation of vulnerabilities. While there are a limited number of these types of systems, they do exist. TestREx \cite{RN110} is one of the most comparable works to this project. The system is intended to provide tooling to systematically test exploits against web applications in a controlled and reproducible environment. However, the architecture and container framework are designed to run vulnerabilities and exploits all in a single container. This type of design limits the telemetry, attack realism, and container reuse provided by SETC. Another similar system is BugBox \cite{RN132}. While BugBox provides a framework that can perform automated end-to-end vulnerability testing, it is explicitly built for deploying and testing PHP applications.

As SETC provides a feature-rich offensive security capability, the framework is also closely related to some automated penetration testing research. One such system, Vapebridge \cite{RN57}, combines scanning, result analysis, exploitation, and reporting. Like Vapebridge, SETC can remotely control exploitation tools to target specific vulnerabilities deployed in container environments. SETC was also designed to allow for fully automated exploitation through scanning and exploitation as proposed in systems such as reinforced learning frameworks \cite{RN55} and IAPTS \cite{RN56}. Currently, fully autonomous vulnerability discovery and exploitation are underdeveloped but influence SETC's design.

The final type of work that was influential and related to SETC design was systems to automate the deployment of cyber ranges and security test environments. Labtainers \cite{RN107, RN105} is a system for creating monitored security environments. While its intended purpose mainly involves deploying security exercises for educational purposes, its auditing and monitoring capacities were influential for SETC. While cyber ranges are not designed to deploy end-to-end automated vulnerability exploitation, they are related to deploying preconfigured environments for security exercises. Two notable cyber range deployment frameworks include Crack \cite{RN108} and CRATE \cite{RN134}. The approaches these systems utilized were examined when designing SETC. However, as they only focus on environment deployment capabilities, the system use cases and capabilities significantly differ from SETC. 

\section{Conclusion}\label{conclusion}

This research presented SETC, a novel framework that automates customizable exploit orchestration and telemetry collection at scale. SETC addresses core challenges in generating rich, diverse data on software vulnerabilities and their exploits. The framework innovates through its configurable and extensible architecture to define modular exploits, vulnerable services, monitoring, and structured logging pipelines.

Compared to current manual processes, SETC enables automated, repeatable, and tailored vulnerability testing to produce extensive telemetry. The research contributes a platform to overcome key barriers in exploit data generation for robust defensive security research. Through orchestrating the end-to-end vulnerability analysis process, SETC can provide diverse, structured data to drive innovations in threat modeling, detection techniques, analysis, and remediation strategies.

The capabilities of SETC were demonstrated through an example scenario which showcases how the framework can enable experiments into logging standards, detection, incident response, and other areas. With its advances in customizable exploit orchestration and telemetry collection, SETC represents a valuable new platform to support impactful vulnerability and defensive security research.

\section{Availability}\label{avail}
The SETC framework aims to advance security research, tooling, and telemetry collection. The framework is fully open source and available under the MIT license. The proof of concept version used for this research, along with future releases, can be found at \url{https://github.com/hackgnar/setc}.

\balance
\printbibliography

\end{document}